# CHARACTERIZING MATERIAL EFFECTS ON DIRECT TOF SIGNAL RESPONSE IN OPTICAL TACTILE SYSTEMS

I. Aulika*[1], A. Ogurcovs[1], M. Kemere[1], A. Bundulis[1], K. Kundzins[1], J. Butikova[1], A. Vembris[1], A. Sarakovskis[1], E. Bacher[2], M. Laurenzis[2], S. Schertzer[2], J. Stopar[3], A. Zore[3], R. Kamnik[3], M. Munih[3], S. Cartiel[4], J. Garcia-Pueyo[4], A. Muñoz[4]

[1]Institute of Solid State Physics, University of Latvia, Riga, Latvia
[2]French-German Research Institute of Saint-Louis, SAINT-LOUIS Cedex, France
[3]University of Ljubljana, Faculty of Electrical Engineering, Ljubljana, Slovenia
[4]University of Zaragoza, Zaragoza, Spain
*ilze.aulika@cfi.lu.lv

**ABSTRACT.** Optical tactile sensing holds transformative potential for robotics, particularly in collaborative environments where touch perception enhances safety, adaptability, and cognitive interaction. However, traditional tactile technologies based on total internal reflection (TIR) and frustrated total internal reflection (FTIR) - such as those used in touchscreen systems - face significant limitations. These include reliance on multiple infrared light sources and cameras, as well as poor adaptability to the complex, curved geometries often found in robotic systems. To address these challenges, we recently introduced OptoSkin [1], an advanced optical tactile sensor based on direct Time-of-Flight (ToF) technology, enabling touch and pressure detection. In this study, we investigate how specific material properties, particularly light scattering, influence the sensitivity of contact point detection under direct ToF sensing. Four materials with distinct scattering coefficients ranging from 0.02 cm$^{-1}$ to 1.1 cm$^{-1}$ at 940 nm were selected to assess their impact on signal quality across different contact scenarios involving various target surfaces.

**INTRODUCTION.** Optical tactile sensing is emerging as a key technology in robotics, human-machine interaction, and wearable systems, offering high-resolution, non-invasive touch and proximity detection. Among the many approaches explored, FTIR has been widely adopted due to its robustness in traditional touchscreen applications. However, FTIR faces critical limitations - such as sensitivity to surface curvature, reliance on precise infrared alignment, and latency issues - which hinder its effectiveness in flexible, dynamic, or non-planar robotic environments. To address these challenges, alternative methods like direct ToF LiDAR-based optical sensing have gained attention. These systems provide faster response times, greater design flexibility, and reduced complexity, making them better suited for modern tactile sensing tasks. However, the performance of ToF-based sensors is highly dependent on the optical properties of the interface material. For example, materials with high transparency and low scattering improve touch accuracy but might be sensitive on contact zone quality. This study introduces some investigations how material properties - specifically diffuse reflectance and scattering coefficient - affect the performance of ToF-based touch sensing. Using a diverse set of materials, we analyse their impact on signal-to-noise ratio (SNR) of ToF.

**EXPERIMENTAL.** A total of four samples were analyzed in this study to evaluate SNR of ToF. One sample was fabricated using a Formlabs 3D printer with Formlabs RS-F2-GPCL-04 clear resin. After printing, the sample underwent UV post-curing at 60°C to ensure structural stability and surface hardness. Additionally, a multilayer variant of the Formlabs sample was produced using a spin-coating process. The first layer was applied at 300 rpm for uniformity and cured for 30 seconds. Subsequent layers were spin-coated at 150 rpm, each also cured for 30 seconds using a 365 nm UV diode. The thickness of individual layers was estimated at 120-150 μm. A third sample was printed using Liqcreate Clear Impact resin with the Elegoo Saturn 2 3D printer. The fourth sample, acrylic glass, was a commercially sourced, optically transparent material.

The optical properties of each material, including scattering and reflectance, were characterized using a Cary 7000 spectrophotometer, equipped with a universal measurement accessory and a diffuse reflectance accessory. This characterization followed the methodological framework developed by F. Foschum et al. [2, 3], which enables precise quantification of scattering behavior in polymeric media.
To evaluate ToF performance, the AMS-OSRAM TMF8828 sensor was used for all direct ToF measurements. The sensor communicated via the I²C protocol, with data acquisition managed by an ESP32-S3 microcontroller, capable of interfacing with up to four sensors simultaneously. For more details, please see our previously published paper [1].
The first experimental configuration involved placing a silicone test object in direct contact with each sample surface, at a fixed distance of 30 mm from the ToF sensor. To assess the impact of other target material impact on contact point, additional plastic test objects (20×20 mm, glossy finish) in white, gray, and black colors were positioned identically. These tests aimed to detect signal variations caused by light leakage or internal scattering, particularly in additively manufactured materials.

**RESULTS.** The scattering coefficient and diffuse reflectance of four tested materials are summarized in the Table 1. The Figure 1 presents a series of 3D color maps illustrating the relationship between diffuse reflectance, scattering coefficient, and the logarithm of the signal-to-noise ratio ($\log_{10}(SNR)$) for four different ToF target interactions with studied materials. Each subplot - (a) silicone, (b) white plastic, (c) gray plastic, and (d) black plastic - represents a different type of contacting or proximate surface, while the color maps visualize how material optical properties influence the resulting SNR under direct ToF sensing. In all subplots, the x-axis denotes diffuse reflectance, the y-axis represents the scattering coefficient, and the color gradient encodes $\log_{10}(SNR)$. Warmer colors (e.g., red and yellow) correspond to higher SNR values, indicating better detection performance, while cooler colors (blue shades) indicate reduced signal quality. The experimental material samples (A-D) are positioned within the maps to show their specific reflectance-scattering combinations and their resulting SNR behavior for each target.

**Table 1.** Information on video and audio files that can accompany a manuscript submission.

| Sample | Material | Scattering coefficient, cm$^{-1}$ | Diffuse reflectance, % |
|---|---|---|---|
| A | Acrylic glass | 0.02 | 0.17 |
| B | FormLabs Clear – Multi layer | 0.1 | 1.57 |
| C | FormLabs Clear – Single layer | 0.2 | 4.46 |
| D | Liqcreate - Clear Impact | 1.1 | 7 |

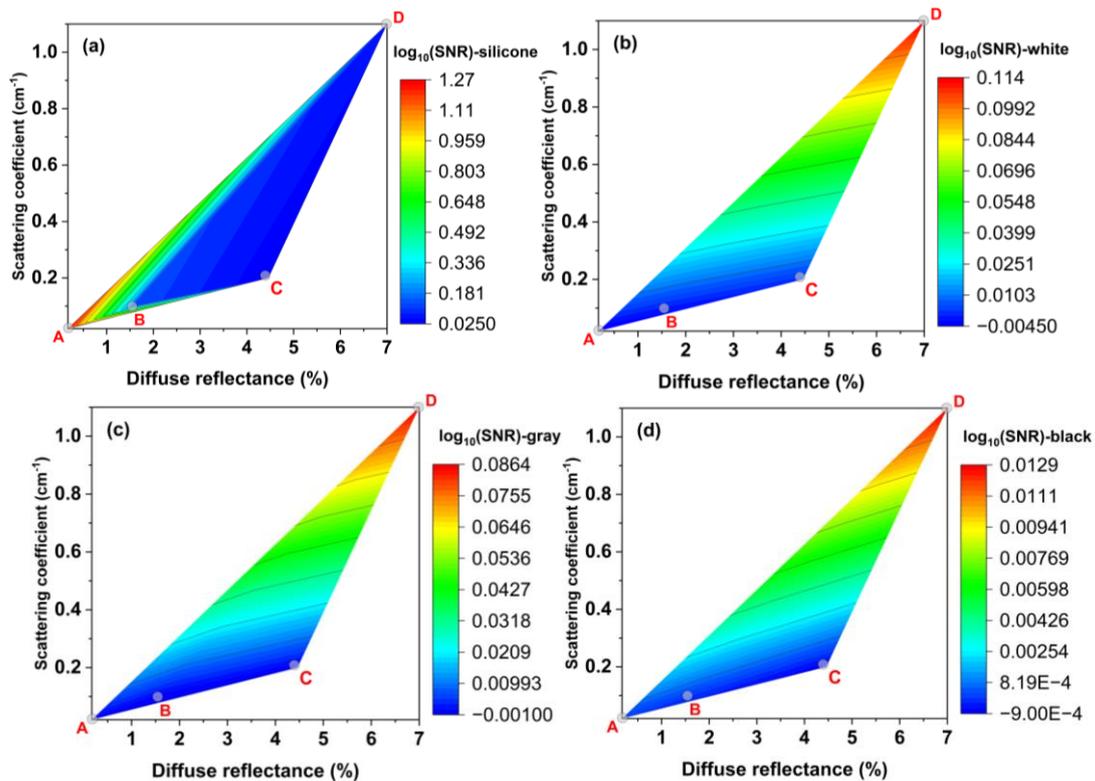

**Figure 1.** 3D color maps showing the relationship between diffuse reflectance, scattering coefficient, and $\log_{10}(SNR)$ for four materials in contact with different targets: (a) silicone, (b) white plastic, (c) gray plastic, and (d) black plastic. Note: the color scale is independently adjusted for each subplot to optimize visual contrast and interpretation.

The results reveal that material performance in ToF sensing strongly depends on the interaction between optical properties and the surface characteristics of the target. For example, acrylic glass (Sample A), with minimal scattering and diffuse reflectance, consistently yields high SNR for silicone contact (Fig. 1a), but performs less effectively with white and gray plastic targets (Fig. 1bc), likely due to poor optical coupling and surface contact quality with the targets. In contrast, Liqcreate Clear Impact (Sample D), which combines high scattering and reflectance, achieves improved performance with white and gray plastics but shows a reduced SNR with the silicone target. This trend highlights the trade-off between clarity and scattering in balancing tactile vs near-field sensitivity.

A more comprehensive overview of this study, including an expanded analysis with a broader range of materials, will be soon available in the MDPI journal Materials [4].

**CONCLUSION.** This study demonstrates that the optical properties of light-guiding materials - specifically diffuse reflectance and scattering coefficient - play a critical role in the performance of direct ToF tactile sensing. Materials with low scattering and reflectance, such as acrylic glass, are well-suited for direct contact sensing with transparent or soft targets like silicone. In contrast, materials exhibiting higher scattering and reflectance, such as Liqcreate Clear Impact, perform better in detecting contact with targets that provide poor optical coupling, particularly opaque materials like white or gray plastics. These findings underscore the importance of selecting light-guiding materials with tailored optical characteristics to optimize both contact and near-field detection in advanced optical tactile systems.

**ACKNOWLEDGEMENTS**. This project is granted from the European Commission's HORIZON EUROPE Research and Innovation Actions under GA number 101070310 References.

**NOTE.** This is the preprint version of the article accepted for publication in SPIE Proceedings: Aulika I. et al., Proc. SPIE 13527, 1352705 (2025), https://doi.org/10.1117/12.3056377.

**References**
[1]   E. Bacher, S. Cartiel, J. García-Pueyo, J. Stopar, A. Zore, R. Kamnik, I. Aulika, A. Ogurcovs, J. Grube, A. Bundulis, J. Butikova, M. Kemere, A. Munoz, and M. Laurenzis, "OptoSkin: Novel LIDAR Touch Sensors for Detection of Touch and Pressure within Wave guides," *IEEE Sensors Journal* **24**(20), 33268 (2024). https://doi.org/10.1109/JSEN.2024.3443615
[2]   F. Foschum, F. Bergmann, and A. Kienle, "Precise determination of the optical properties of turbid media using an optimized integrating sphere and advanced Monte Carlo simulations. Part 1: theory," *Appl. Opt.* **59**(10), 3203–3215 (2020). https://doi.org/10.1364/AO.386011
[3]   F. Bergmann, F. Foschum, Ralf Zuber, and A. Kienle, "Precise determination of the optical properties of turbid media using an optimized integrating sphere and advanced Monte Carlo simulations. Part 2: experiments," *Appl. Opt.* **59**(10), 3216–3226 (2020). https://doi.org/10.1364/AO.385939
[4]   I. Aulika, A. Ogurcovs, M. Kemere, A. Bundulis, K. Kundzins, J. Butikova, A. Vembris, A. Sarakovskis, E. Bacher, M. Laurenzis, S. Schertzer, J. Stopar, A. Zore, R. Kamnik, M. Munih, S. Cartiel, J. García-Pueyo, and A. Munoz, "Influence of Material Optical Properties in Direct ToF LiDAR Optical Tactile Sensing," *to be submitted to Materials (MDPI)* (2025).